\documentclass{appolb}
\usepackage{epsfig}
\def\be{\begin{equation}}
\def\ee{\end{equation}}
\def\bea{\begin{eqnarray}}
\def\eea{\end{eqnarray}}

\begin{document}
\title{Improved description of Bose-Einstein Correlation function%
\thanks{Presented by T. Nov\'ak at the $4^{\mathrm{th}}$ 
Budapest Winter School on Heavy Ion Physics, December 1 - 3, 2004. To be published
in Acta Physica Hungarica Heavy Ion Physics.}%
}
\author{T. Nov\'ak \\ Representing L3 Collaboration
\address{Institute for Mathematics, Astrophysics and Particle Physics, \\
Radboud University Nijmegen 6525 ED Nijmegen, The Netherlands\\[0.2ex] 
}
}
\maketitle
\begin{abstract}
The L3 data on Bose-Einstein correlations of equally charged 
  pion pairs produced in hadronic Z decays are analyzed in terms of
  various parametrizations. Preliminary results are presented here.
\end{abstract}
  
\section{Introduction}
In particle and nuclear physics intensity interferometry provides a direct
experimental method for the determination of sizes, shapes and lifetimes
of particle emitting sources (for recent reviews see \cite{bib1,bib2}).
In particular, boson interferometry provides a powerful tool for the 
investigation of the space-time structure of particle production processes, 
since Bose-Einstein correlations (BEC) of two identical bosons reflect both 
geometrical and dynamical properties of the particle radiating source.

Originally, the method of Bose-Einstein correlations was invented by the 
radio astronomers R. Hanbury Brown and R. Q. Twiss (HBT), who applied it to
determine the angular diameters of main sequence stars \cite{bib3,bib4}.
The first experimental evidence for Bose-Einstein correlations in high 
energy physics dates back to 1959 \cite{bib5}. The results were interpreted 
as BEC by G. and S. Goldhaber, Lee and Pais (GGLP) \cite{bib6}. Angular 
distributions of pions could be described more exactly by applying 
Bose-Einstein statistics instead of a classical statistical model. This effect 
is frequently referred as either the HBT, or GGLP effect, or simply 
Bose-Einstein correlations. 

The L3 experiment provides a good opportunity to perform detailed investigation
of Bose-Einstein correlations in $\mathrm{e}^+\mathrm{e}^-$ 
annihilation at a center of mass 
energy of $\sqrt{s}=91$ GeV. After ``standard'' event and track selection 
we concentrate on 2-jet events detected using the Durham algorithm \cite{bib7}
with resolution parameter $y_{\mathrm{cut}}=0.006$. Finally, approximately 13 million 
pion pairs are analyzed. The results presented here are preliminary.

\section{Shape of the BEC function}\label{techno}  
The two-particle Bose-Einstein correlation function is defined as:
\begin{equation}
  C_2(p_1,p_2) = \frac{\rho_2(p_1,p_2)}
    {\rho_1(p_1) \rho_1(p_2)},
\end{equation}
where $\rho_2(p_1,p_2)$ is the two-particle invariant momentum 
distribution, $\rho_1(p_i)$ is the single-particle invariant momentum 
distribution and $p_i$ stands for the four-momentum of particle $i$. Since 
it is difficult to create 
the product of the single-particle distribution it is replaced
by a ``reference sample'', the two-particle density that would occur
in the absence of BE interference.
 
If long-range correlations can be neglected or corrected for, and if the 
short-range correlations are dominated by Bose-Einstein correlations,
this two-particle BEC function is related to the Fourier-transformed
source distribution. If we assume that $f(x)$ is the density distribution
of the source of the pions then the correlation function 
is found to be
\begin{equation}
  C_2(p_1,p_2) = 1 + |\tilde{f} (Q)|^2 ,
\end{equation}
where $Q$ is the invariant four-momentum difference, $Q=-(p_1-p_2)^2$ and
$\tilde{f} (Q)$ is the Fourier transform of $f(x)$
\begin{equation}
  \tilde{f} (Q) = \int \mathrm{d} x \exp (iQx) f(x).
\end{equation}
 
\subsection{Gaussian distributed source}\label{details}
The simplest assumption is that the source has a Gaussian distribution,
in which case the Fourier transformed source function is determined as
$\tilde{f} (Q)=\exp \left(-\frac{Q^2}{2}\right)$
which yields
\begin{equation}\label{gauss}
  C_2(Q) = \gamma \left[1 + \lambda \exp \left(-R^2Q^2 \right) \right] 
           \left( 1 + \delta Q \right),
\end{equation}
where the parameter $\gamma$ is a constant of normalization,
$\lambda$ is an incoherence factor, which measures the strength of
the correlation, $R$ is a scale parameter which measures the length of
homogeneity and $\left( 1 + \delta Q \right)$ is introduced to parametrize
possible long-range correlations.

\begin{figure}[!thb]
\begin{center}
\vspace{-0.5cm}
\includegraphics[angle=0,width=0.7\textwidth]{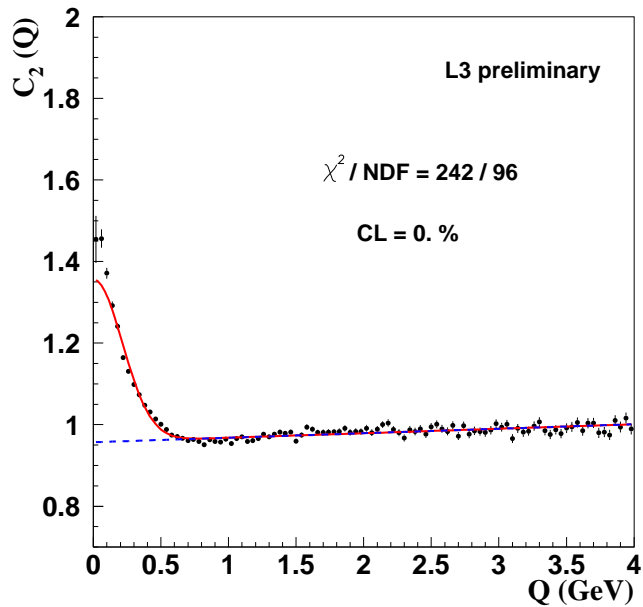}
\vspace{-0.5cm}
\end{center}
\caption[]{The Bose-Einstein correlation function $C_2$. The full line 
corresponds to the fit by Eq.\ref{gauss}, while the dashed line stands for the 
long-range momentum correlations.

}
\label{gaussfig}
\end{figure}

A fit of this correlation function results in an unacceptably low confidence
level from which one concludes that the shape of the source deviates from a 
Gaussian. The fit is particularly bad at low $Q$ values, as shown in Fig. 
\ref{gaussfig}.

\subsection{L\'evy distributed source}\label{maths}

The Central limit theorem states that, under certain conditions, the sum 
of a large 
number of random variables behaves as a Gaussian distribution. The 
generalization of the central limit theorem gives the classification of the 
stable distributions. The study of these stable distributions was begun 
by Paul L\'evy in the 1920's. For recent results see \cite{bib8}. 

According to the generalized Central Limit Theorem, if particle production is 
a result of a multifold probabilistic process such as the branching of 
gluons into gluons into gluons etc., then the class of possible limiting 
probability distributions coincides with the class of L\'evy 
distributions \cite{bib8}. The characteristic function of symmetric 
stable distributions is \cite{bib9}
\begin{equation}
  \tilde{f}(Q) = \exp \left(iQ\delta - \frac{ \left| R Q \right| ^\alpha}{2}\right).
\end{equation}
Here we utilize the notation of Nolan \cite{bib8}. The index of stability,
$\alpha$, satisfies the inequality $0<\alpha \leq 2$. The case $\alpha=2$
corresponds to a Gaussian source distribution. For more details see 
\cite{bib8}.

\begin{figure}[!thb]
\begin{center}
\vspace{-0.5cm}
\includegraphics[angle=0,width=0.7\textwidth]{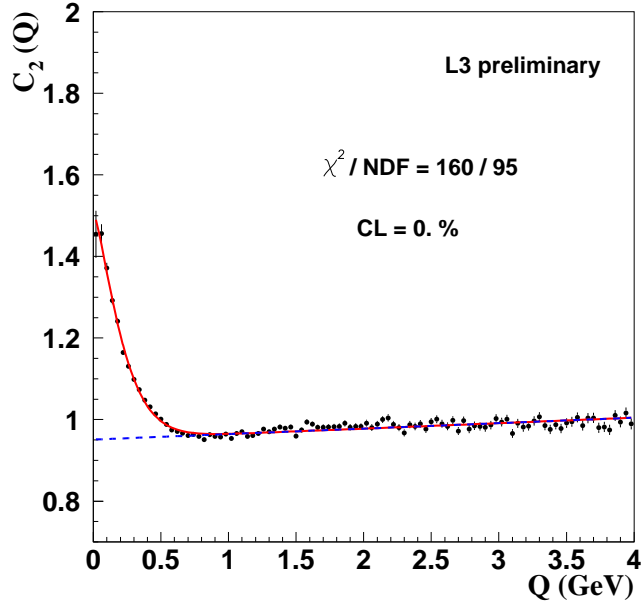}
\vspace{-0.5cm}
\end{center}
\caption[]{The Bose-Einstein correlation function $C_2$. The full line 
corresponds to the fit by Eq.\ref{gauss}, while the dashed line stands for the 
long-range momentum correlations.

}
\label{fig1}
\end{figure}

Thus the Bose-Einstein correlation function for L\'evy stable distributions 
has the following, relatively simple form:
\begin{equation}\label{symlev}
  C_2(Q) = \gamma \left[ 1+ \lambda \exp \left(-(RQ)^\alpha \right) \right]
          (1+ \delta Q)
\end{equation}
After fitting the data with Eq.\ref{symlev} it is clear that the correlation
function is far from Gaussian: $\alpha \approx 1.3$. However, the confidence 
level is still unacceptably low. The curve of this fit is shown
in Fig. \ref{fig1}.

Since there is no particle production before the onset of the 
collision, a more appropriate form of the source distribution
for the time component is the  asymmetric stable distribution. In this 
case, one obtains the following result for the correlation function:
\begin{equation}\label{asymlev}
  C_2(Q) = \gamma \left[ 1+ \lambda \cos \left[(R_aQ)^ \alpha \right]
           \exp \left(-(RQ)^\alpha \right) \right] (1+ \delta Q),
\end{equation}
where $R_a$ is an additional parameter.
\begin{figure}[!thb]
\begin{center}
\vspace{-0.5cm}
\includegraphics[angle=0,width=0.7\textwidth]{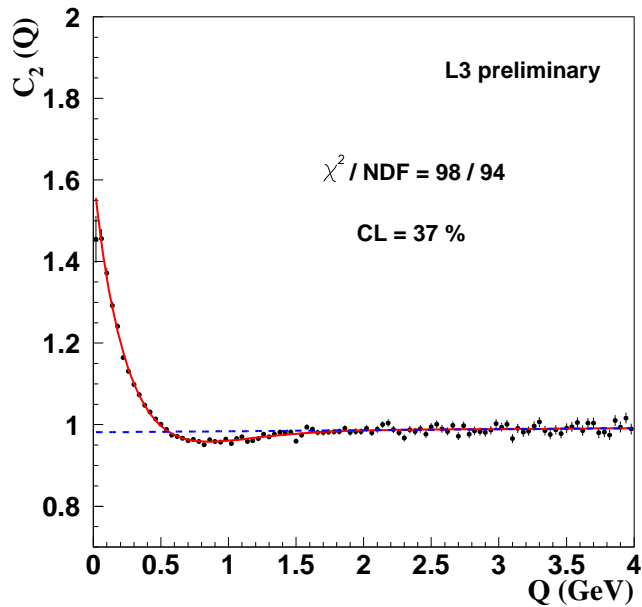}
\vspace{-0.5cm}
\end{center}
\caption[]{The Bose-Einstein correlation function $C_2$. The full line 
corresponds to the fit by Eq.\ref{gauss}, while the dashed line stands for the 
long-range momentum correlations.

}
\label{fig2}
\end{figure}

The fit of Eq.\ref{asymlev} to the data is statistically acceptable. The data 
points are well described by the fit curve which is 
shown in Fig. \ref{fig2}. Note that for $Q$ between 
0.5 $\mathrm{GeV}$ and 1.5 $\mathrm{GeV}$ the fitted curve goes below one. 
This is caused by the cosine term, which comes from the asymmetric L\'evy 
assumption, in Eq.\ref{asymlev}.
The fitted value of the index of stability, 
$\alpha$ is found to be approximately 0.8.
 
\section{Conclusions}\label{others}
The assumption that the source has a Gaussian shape is too simple.
A good description of the Bose-Einstein correlation function is 
achieved using L\'evy stable distributions as the source function.
 
\section*{Acknowledgments}
The author would like to thank Tam\'as Cs\"org\H{o}, Wolfram Kittel and 
Wes Metzger for inspiration, support and careful attention. He is also
grateful to the members of the L3 collaboration.

\end{document}